\documentclass[conference, a4paper]{IEEEtran}
    \newtheorem{theorem}{Theorem}
    \newtheorem{lemma}[theorem]{Lemma}
    \newtheorem{proposal}[theorem]{Proposal}
    
    \newtheorem{definition}[theorem]{Definition}
    \newtheorem{conjecture}[theorem]{Conjecture}
    \newtheorem{proposition}[theorem]{Proposition}

\usepackage[cmex10]{amsmath} 
\usepackage{mathtools, amssymb, upgreek, mleftright}
    \def\bb{\upbeta}
    \def\CC{{\mathbf C}}
    \def\>{\succcurlyeq}
    \def\<{\preccurlyeq}
    \def\ato{\mathrm{@0}}
    \def\atl{\mathrm{@1}}
    \def\avg{\mathrm{avg}}
    \def\fst{\mathrm{fst}}
    \def\hlf{\mathrm{hlf}}
    \def\Ber#1{{\mathrm{Ber}(#1)}}
    
\usepackage{tikz-cd, booktabs, caption, subcaption} 
    \tikzset{every picture/.style={cap=round, join=round}}

\usepackage[noadjust]{cite} 
\usepackage{xurl, hyperref}
    \hypersetup{
        colorlinks, allcolors={rgb,"FF: red,"00; green,"11; blue,"99},
        pdfsubject={Information Theory (cs.IT)},
        pdfkeywords={
            polar code,
            partial order,
            beta expansion,
            Bernstein basis,
            hard threshold,
            scaling exponent,
        },
    }

\begin{document}

\advance\baselineskip0pt plus.1pt minus.05pt
\advance\lineskip0pt plus.1pt minus.05pt
\advance\parskip0pt plus.2pt minus.1pt

\title{Fast Methods for Ranking Synthetic BECs}

\author{%
    \IEEEauthorblockN{Hsin-Po Wang}%
	\IEEEauthorblockA{%
		University of California, Berkeley, CA, USA\\
		simple@berkeley.edu%
	}%
    \and
    \IEEEauthorblockN{Vlad-Florin Drăgoi}%
    \IEEEauthorblockA{%
        Aurel Vlaicu University of Arad, Romania\\
        LITIS, University of Rouen Normandie, France\\
        vlad.dragoi@uav.ro%
    }%
}

\maketitle

\begin{abstract}
    We gather existing methods that are used to compare and rank the
    BECs synthesized by a polar code constructor, compare them, and
    propose new methods that compare synthetic BECs faster.
\end{abstract}

\section{Introduction}

    The discovery of polar coding enabled 5G and other coding tasks.
    The formalism introduced by Arıkan \cite{Ari09} came with a large
    panel of theoretical and practical challenges.  A fundamental one is
    about the reliability of the synthetic channels $W^\alpha$, where
    $W$ is the underlying physical channel and $\alpha$ is a binary
    string, and how to order the set $\{W^\alpha : \alpha \in
    \{0,1\}^m\}$.  Further, we shall write $\alpha \> \gamma$ when
    $W^\alpha$ is favored over $W^\gamma$ for all $W$.

    From general channel models ($W$ being a BMS channel) to particular
    ones ($W$ being a BEC), several rules were proposed.  Rule Set A
    (RS-A) was initially used by Mori and Tanaka \cite{MoT09c} to
    construct polar codes over BMS channels and is generated by $1 \>
    0$.  Later, Schurch \cite{Sch16} and Bardet et al.\ \cite{BDO16}
    added $10\> 01$ over BMS channels, which we call Rule Set B (RS-B).
    
    Other works focus on BECs.  Dragoi and Cristescu \cite{DrC21}
    introduced a family of infinitely many new rules: $10 0^k 01 \> 01
    0^k 10$ for any integer $k \geq 0$; they are dubbed Rule Set C
    (RS-C).  A recent work \cite{Conjugate} generalizes RS-C to even
    more rules that are not cataloged here.  Wu and Siegel \cite{WuS19}
    conjectured that $0^k 1^{2^k} \> 1^k 0^{2^k}$ is true for any
    integer $k \geq 0$, which is claimed to be proved in \cite{ODE}; we
    call them Rule Set D (RS-D).  Kahraman \cite{Kah17} proposed that
    the first $2^k$ terms of the Thue--Morse sequence should be inferior
    to their bitwise complement for all integer $k \geq 0$.  This family
    begins with 
    (a) $0 \< 1$;
    (b) $01 \< 10$;
    (c) $0110 \< 1001$;
    (d) $01101001 \< 10010110$.
    Observe that (a), (b), and (c) can be explained by RS-A, RS-B, and
    RS-C, respectively.  We call these Rule Set E (RS-E).

    Order relations for polar codes have several applications.  Bardet
    et al.\ \cite{BDO16} conducted the first investigation of
    automorphism groups of polar codes that respect RS-A and RS-B.  The
    automorphism group then plays a role in parallelized decoding of
    polar codes \cite{ZWT21, GEE21e, GEE21g, PBL21, PBL22, IvU22e,
    IvU22h}.  Another application involves construction of polar codes.
    Mondelli, Hassani, and Urbanke showed in \cite{MHU19} that the
    reliability of all but $o($block length$)$ synthetic channels can be
    inferred using RS-A and RS-B, hence the sub-linearity of the
    complexity of code construction.  (For other approaches of code
    construction, see \cite{OlD22} for discussions on Gaussian
    approximation.)

    As for a better understanding of $W^\alpha$ itself, the sharp
    transition of $I(W^\alpha)$ with respect to $I(W)$ was analyzed in
    \cite{OrR19}.  The position of the threshold was investigated in
    \cite{Gei18}. Alongside, the link between network reliability and
    synthetic channels over BECs was established.  The behaviors for
    $I(W)$ very close to $0$ and $1$ were studied in \cite{DrB20}, while
    the average behavior of $I(W^\alpha)$ we studied in \cite{DrC21}.

    On a parallel track, He et al.\ proposed \emph{beta expansion}
    \cite{HBL17} that bypasses partial order relations but aims directly
    at construction of codes.  The idea is to ``evaluate'' a binary
    string $1001$ to a real number $1001_\bb \coloneqq 1\cdot\bb^3 +
    0\cdot\bb^2 + 0\cdot\bb^1 + 1\cdot\bb^0$ before comparing the
    evaluations.  Simulations over AWGN channels determined that a good
    value of $\bb$ is $2^{1/4}$.

    Eventually, we could retrieve the maximum amount of information
    about the poset using all the techniques and all the
    total/partial/pre-order relations.  Our problem can be restated as
    follows: how to decide whether $\alpha$ and $\gamma$ are comparable
    or not with respect to $\<$ for all pairs $(\alpha, \gamma)$ of
    binary strings with length at most $m$.  We attack this problem from
    three sides: (a) generate as many comparable pairs as possible, (b)
    generate as many non-comparable pairs as possible, and (c) generate
    total orders that (almost) extend the targeted partial order.

\begin{figure*}
    \centering
    \begin{tikzpicture} [y=7mm]
        \tikzset{
            partial/.style={draw, circle},
            total/.style={draw, rectangle},
        }
        \path
            (-8, 0) node [total, scale=2.5] (beta) {$\bb$}
            (-4, 3) node [partial] (A) {RS-A}
            (-4, 1.5) node [partial] (B) {RS-B}
            (-4, 0) node [partial] (C) {RS-C}
            (-4, -1.5) node [partial] (D) {RS-D}
            (-4, -3) node [partial] (E) {RS-E}
            (0, 0) node [partial, scale=3] (std) {$\<$}
            (4, 3.2) node [total] (avg) {$\avg\strut$}
            (4, 2) node [total] (ato) {$\ato\strut$}
            (4, 0) node [partial] (fst) {$\fst\strut$}
            (4, -2) node [total] (atl) {$\atl\strut$}
            (4, -3.2) node [total] (hlf) {$\hlf\strut$}
            (8, 0) node [partial, scale=1.5] (Ber) {$\Ber{n}$}
        ;
        \definecolor{known}{HTML}{0011AA}
        \definecolor{conj}{HTML}{880011}
        \definecolor{our}{HTML}{116600}
        \tikzset{
            every node/.style={auto, sloped},
            every edge/.style={
                draw, shorten <=2pt, shorten >=2pt,
                commutative diagrams/Rightarrow
            }
        }
        \hypersetup{allcolors=.} 
        \makeatletter
            \let\oldinner\pgf@stroke@inner@line
            \def\pgf@stroke@inner@line{\pgfsetroundcap\oldinner}
        \makeatother
        \path [every double/.style={cap=butt}]
            (A) edge [known, bend right=24]
                node {$\bb > 0$ \cite{HBL17}} (beta)
            (B) edge [known, bend right=12]
                node {$\bb > 1$ \cite{HBL17}} (beta)
            (C) edge [our, ']
                node {$\bb > 1$ (Prop~\ref{pro:rsC=>beta})} (beta)
            (D) edge [our, bend left=12, '] node
                {$\bb \leq 2^{-1/k}$ (Prop~\ref{pro:rsD-beta})} (beta)
            (E) edge [our, bend left=24, ']
                node {$\bb > 1$ (Prop~\ref{pro:rsE=>beta})} (beta)
            (A) edge [known, bend left=24] node {\cite{MoT09c}} (std)
            (B) edge [known, bend left=12]
                node {\cite{Sch16, BDO16}} (std)
            (C) edge [known] node {\cite{DrC21, Conjugate}} (std) 
            (D) edge [conj, bend right=12]
                node {\cite{WuS19}, \color{known}\cite{ODE}} (std)
            (D) edge [known, bend right=12, '] node {\cite{WuS19}} (hlf)
            (E) edge [our, bend right=24, ']
                node {\cite{Kah17} \& Prop~\ref{pro:rsE}} (std)
            (std) edge [known, bend left=24] node {\cite{DrC21}} (avg)
            (std) edge [our, bend left=16]
                node {Prop~\ref{pro:std=>at}} (ato)
            (std) edge [our] node [yshift=-5pt]
                {Prop~\ref{pro:std=>fst}} (fst)
            (std) edge [our, bend right=16, ']
                node {Prop~\ref{pro:std=>at}} (atl)
            (std) edge [known, bend right=24, ']
                node {\cite{WuS19}} (hlf)
            (std) edge [conj, bend left=12]
                node {conditional (Prop~\ref{thm:std=>Ber})} (Ber)
            (Ber) edge [our, bend left=12, ']
                node {any $n$ (Prop~\ref{pro:Ber=>std})} (std)
            (Ber) edge [our, bend right=24] (avg)
            (Ber) edge [our, bend right=16] (ato)
            (Ber) edge [our] (fst)
            (Ber) edge [our, bend left=16] (atl)
            (Ber) edge [our, bend left=24] (hlf)
        ;
    \end{tikzpicture}
    \caption{
        Ordering relations for polar codes.  Inspired by
        \cite[Figure~5]{DrC21}.  Rectangles are total preorders.
        Circles are partial preorders.  Arrows are implications, e.g.,
        $\alpha \< \gamma$ implies $\alpha \leq_\avg \gamma$.  Blue
        for existing works; green for ours; red for conjectures.  The
        arrow from RS-D to $\<$ is claimed to be proved in \cite{ODE}.
    }                                                \label{fig:mitosis}
\end{figure*}
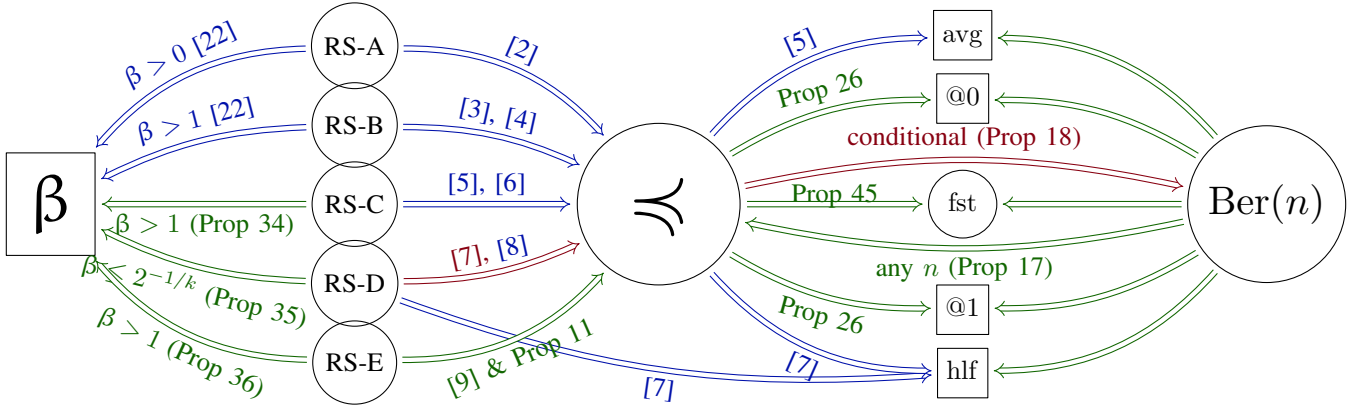

    One unusual thing we do is not to restrict the scope to comparing
    bit strings of the same length, as most of the existing works did.
    Instead, we study \emph{intergenerational} comparisons.  Firstly, we
    observe that some intragenerational inequalities are consequences of
    intergenerational ones; an example is that $011 \> 10$ leads to
    $0011 \> 010 \> 1000$.  Secondly, intergenerational inequalities
    compares synthetic channels for relaxed polar codes \cite{EMF17},
    pruned polar codes \cite{WaD21}, and parallelized low-latency polar
    codes \cite{HMF22}.

    While this paper is more of a survey than a research paper, we still
    make several contributions along the way:
    \begin{itemize}
        \item We show that RS-E is true over BECs
        (Proposition~\ref{pro:rsE}).
        \item We show that RS-C and RS-E are compatible with beta
        expansion.  But RS-D becomes not compatible with beta expansion
        as the string length approaches infinity (Propositions
        \ref{pro:rsC=>beta} to \ref{pro:rsE=>beta}).
        \item We propose a total preorder, $\>_\fst$, that is easy to
        compute, compatible with all rule sets, and only slightly
        different from $\>$ (Section~\ref{sec:fast}).
        \item We present evidence that the most appropriate $\bb$ value
        is related to the scaling exponent of polar codes
        (Section~\ref{sec:scale}).
    \end{itemize}

    This paper is organized as follows.
    Section~\ref{sec:poset} defines the reliability poset.
    Section~\ref{sec:rs} goes over the rule sets.
    Section~\ref{sec:Ber} discusses Bernstein basis.
    Section~\ref{sec:beta} discusses beta expansion.
    Section~\ref{sec:cut} studies threshold behavior.

\section{The Reliability Poset of Synthetic BECs}     \label{sec:poset}

    Given a BEC $W$ with capacity $I(W) = x \in [0, 1]$, the polar
    transformation synthesizes two channels, $W^0$ and $W^1$; they  are
    BECs with capacities
    \[
        I(W^0) = I_0(x) \coloneqq x^2
        \quad\text{and}\quad
        I(W^1) = I_1(x) \coloneqq 1 - (1 - x)^2,
    \]
    respectively.  To study the recursion of polar transformations, we
    define $I(x) \coloneqq x$ and
    \[
        I_{a_1 a_2 \dotsm a_\ell}(x)
        \coloneqq I_{a_2 \dotsm a_\ell} (I_{a_1}(x))
    \]
    for any bit string $a_1 a_2 \dotsm a_\ell \in \{0, 1\}^\ell$.  This
    way, $I_\alpha(x)$ is the capacity of $W^\alpha$ for any $\alpha \in
    \{0, 1\}^\ell$.  We call it the \emph{reliability polynomial} of
    $\alpha$.

    We want to study if $W^\alpha$ is no worse than $W^\gamma$ for any
    BEC $W$, and this for any pair of bit strings $\alpha$ and $\gamma$.
    We define a binary relation $\>$.

    \begin{definition}
        For any $\alpha \in \{0, 1\}^\ell$ and $\gamma \in \{0, 1\}^m$,
        where $\ell$ and $m$ are nonnegative integers, we write $\alpha
        \> \gamma$ and say $\alpha$ \emph{outperforms} $\gamma$ if \[
        I_\alpha(x) \geq I_\gamma(x) \quad \text{ for all } x \in [0,
        1]. \]
    \end{definition}

    Many works have studied $\>$.  So one might be surprised when
    learning that we are the first to prove the following.

    \begin{theorem}
        $\>$ is a partial order on $\bigcup_{m=0}^{\infty} \{0, 1\}^m$.
    \end{theorem}

    \begin{IEEEproof}
        Reflexivity: since $I_\alpha(x) \geq I_\alpha(x)$, $\>$ is
        reflexive.

        Transitivity: since $I_\alpha(x) \geq I_\gamma(x)$ and
        $I_\gamma(x) \geq I_\delta(x)$ implies $I_\alpha(x) \geq
        I_\delta(x)$, $\>$ is transitive.

        Antisymmetry: this is the nontrivial part.  Suppose $\alpha \>
        \gamma \> \alpha$, then $I_\alpha(x) \geq I_\gamma(x) \geq
        I_\alpha(x)$, which implies $I_\alpha(x) - I_\gamma(x) = 0$ for
        all $x \in [0, 1]$.  By the fundamental theorem of algebra, any
        nonzero polynomial has finitely many roots; but $I_\alpha -
        I_\gamma$ has infinitely many so it must be the zero polynomial.
        Now proving antisymmetry boils down to proving $\alpha = \gamma$
        given $I_\alpha = I_\gamma$.  We prove it as a standalone lemma,
        Lemma~\ref{lem:injective}.
    \end{IEEEproof}

    \begin{lemma}                                 \label{lem:injective}
        If $I_\alpha = I_\gamma$ as polynomials, $\alpha = \gamma$ as
        bit strings.
    \end{lemma}

    \begin{IEEEproof}
        We will prove the contrapositive.  Suppose $\alpha =
        \kappa0\lambda$ and $\gamma = \kappa1\mu$.  That is, they both
        begin with $\kappa$ but disagree afterward.  By the fundamental
        theorem of algebra, there exists a complex number $z \in \CC$
        such that $I_\kappa(z) = -1$.  Now $I_\alpha(z) = I_{0\lambda}
        (I_\kappa(z)) = I_{0\lambda}(-1) = I_\lambda (I_0(-1)) =
        I_\lambda(1) = 1$.  On the other hand, $I_\gamma(z) = I_{1\mu}
        (I_\kappa(z)) = I_{1\mu}(-1) = I_\mu (I_1(-1)) = I_\mu(-3)$.
        Note that $|I_0(x)|$ and $ |I_1(x)|$ will be $\geq 3$ if $|x|
        \geq 3$.  So it is impossible that $I_\mu$ will send $-3$ to
        $1$.  This implies $I_\gamma(z) \neq 1 = I_\alpha(z)$; they are
        two different polynomials.
    \end{IEEEproof}

\begin{figure*}
    \centering
    \begin{subfigure}[c]{0.195\textwidth}
        \includegraphics[width=0.999\textwidth]{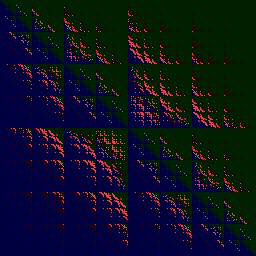}
        \caption{All rule sets}
    \end{subfigure}
    \hfill
    \begin{subfigure}[c]{0.195\textwidth}
        \includegraphics[width=0.999\textwidth]{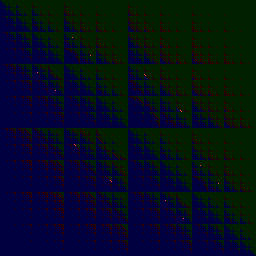}
        \caption{Bernstein ($\Ber{256}$)}
    \end{subfigure}
    \hfill
    \begin{subfigure}[c]{0.195\textwidth}
        \includegraphics[width=0.999\textwidth]{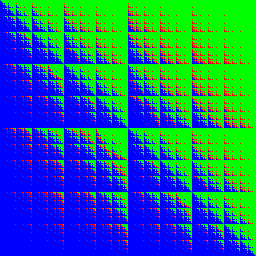}
        \caption{Standard comparison}
    \end{subfigure}
    \hfill
    \begin{subfigure}[c]{0.195\textwidth}
        \includegraphics[width=0.999\textwidth]{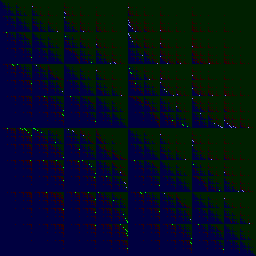}
        \caption{Fast preorder}
    \end{subfigure}
    \hfill
    \begin{subfigure}[c]{0.195\textwidth}
        \includegraphics[width=0.999\textwidth]{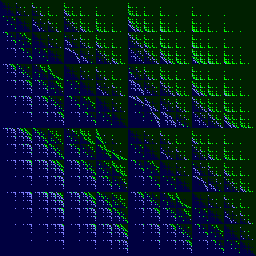}
        \caption{Beta expansion}
    \end{subfigure}%
    \caption{
        Binary relations visualized via incidence matrices.  Each
        picture has $256$ rows and $256$ columns indexed by $\{0, 1\}^8$
        lexicographically.  Blue means row index $\>$ column index.
        Green means row index $\<$ column index.  Red means
        incomparable.  Dimmed pixels are those that agree with (c).
        Note that (b) contains sixteen red pixels not dimmed.  But as
        the $n$ in $\Ber{n}$ increases, all pixels will eventually agree
        with (c).
    }                                                    \label{fig:std}
\end{figure*}

    See the center of Fig.~\ref{fig:std} for the partial order $\>$ on
    $\{0, 1\}^8$.

\section{Predefined Rule Sets}                            \label{sec:rs}

    By \emph{rules} we mean simple inequalities that can be used to
    generate more inequalities.  One conditional and two unconditional
    rules were known very early and they hold for a more general
    context: BMS channels and channel degradation.

    \begin{proposition} [Concatenation {\cite[Lemma~4.7]{Kor09}}]
                                                      \label{pro:concat}
        If $\alpha \> \gamma$ and $\kappa \> \lambda$,
        then $\alpha \kappa \> \gamma \lambda$.
    \end{proposition}

    \begin{proposition} [Rule Set A \cite{MoT09c}]       \label{pro:rsA}
        $1 \> 0$.
    \end{proposition}

    \begin{proposition} [Rule Set B \cite{Sch16, BDO16}] \label{pro:rsB}
        $10 \> 01$.
    \end{proposition}

    And then there are relations proposed with BECs in mind.
        
    \begin{proposition} [Duality {\cite[Corollary~4]{WuS19}}]
                                                        \label{pro:dual}
        If $\alpha \> \gamma$, then $\bar\alpha \< \bar\gamma$.  Bar
        denotes bitwise complement, e.g., $\overline{1000} = 0111$.
    \end{proposition}

    \begin{proposition} [Rule Set C {\cite[Definition~8]{DrC21},
        \cite[Section VII.B]{Conjugate}}]                \label{pro:rsC}
        $10 0^k 01 \> 01 0^k 10$ for any integer $k \geq 0$.
    \end{proposition}

    \begin{conjecture} [Rule Set D {\cite[(45)]{WuS19}}]
                                                         \label{pro:rsD}
        $0^k 1^{2^k} \> 1^k 0^{2^k}$ for any integer $k \geq 0$.
        (This is claimed to be proved in \cite{ODE}.)
    \end{conjecture}

    \begin{proposal} [Rule Set E \cite{Kah17}]           \label{con:rsE}
        $\tau_1 \dotsm \tau_{2^k} \< \overline{\tau_1 \dotsm
        \tau_{2^k}}$ for any integer $k \geq 0$.  Here, $\tau =
        01101001\dots$ is the Thue--Morse sequence \cite[OEIS:
        A010060]{oeis}: $\tau_n$ is the parity of the number of $1$'s in
        the binary representation of $n$.
    \end{proposal}

    Remark: the significance of Proposal~\ref{con:rsE} is that it
    generalizes RS-A, RS-B, and RS-C in an interesting way but,
    concerning the chaotic nature of $\tau_n$, it is not obvious how to
    prove it.  We succeed in finding a proof.

    \begin{proposition}                                  \label{pro:rsE}
        $\tau_1 \dotsm \tau_{4k} \< \overline{\tau_1 \dotsm \tau_{4k}}$
        for any integer $k \geq 1$.
    \end{proposition}

    \begin{IEEEproof}
        \def\0{o}
        \def\1{\iota}
        Let $\0 \coloneqq 0110$ and $\1 \coloneqq 1001$.  Use brute
        force to verify (a) $\0 \< \1$, (b) $\0\1 \< \1\0$, and (c)
        $\0\1\1 \< \1\0\0$.  Next use the fact that $\tau$ is
        \emph{cube-free} (it does not contain $\1\1\1$) to infer that
        $\tau = \0\1\1~ \0\1~ \0~ \0\1\1~ \0~ \0\1~ \dotsm$ can be
        written as a product of $\0$, $\0\1$, and $\0\1\1$.  Hence (a),
        (b), and (c) are all we need to prove $\tau_1 \dotsm \tau_{4k}
        \< \overline{\tau_1 \dotsm \tau_{4k}}$.
    \end{IEEEproof}

    \begin{proposition} [Rule Set F]                     \label{pro:rsF}
        $011 \> 10$.
    \end{proposition}
    
    \begin{IEEEproof}
        $I_{011}(x) - I_{10}(x) = x^3 (x - 1)^2 (4 + x - 2x^2 - x^3)$;
        which is nonnegative over $x \in [0, 1]$.
    \end{IEEEproof}

\begin{figure*}
    \centering
    \begin{subfigure}[c]{0.195\textwidth}
        \includegraphics[width=0.999\textwidth]{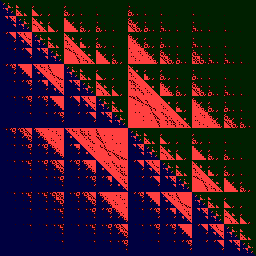}
        \caption{RS-A and B}
    \end{subfigure}
    \hfill
    \begin{subfigure}[c]{0.195\textwidth}
        \includegraphics[width=0.999\textwidth]{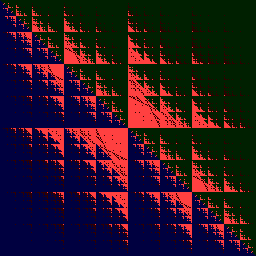}
        \caption{RS-A, B, and C}
    \end{subfigure}
    \hfill
    \begin{subfigure}[c]{0.195\textwidth}
        \includegraphics[width=0.999\textwidth]{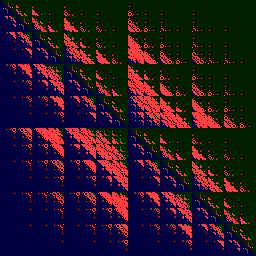}
        \caption{RS-A, B, and D}
    \end{subfigure}
    \hfill
    \begin{subfigure}[c]{0.195\textwidth}
        \includegraphics[width=0.999\textwidth]{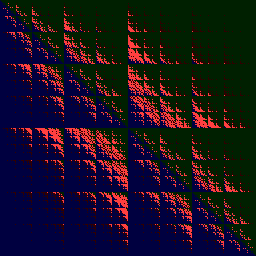}
        \caption{RS-A, B, C, and D}
    \end{subfigure}
    \hfill
    \begin{subfigure}[c]{0.195\textwidth}
        \includegraphics[width=0.999\textwidth]{8rsABCDF.png}
        \caption{RS-A, B, C, D, and F}
    \end{subfigure}%
    \caption{
        Same setting as Fig.~\ref{fig:std}.  As more rule sets are
        considered, less incomparable pairs wander.
    }\label{fig:RS}
\end{figure*}

    See Fig.~\ref{fig:RS} for how progressively adding new rules unlocks
    more comparable pairs.  Notice how RS-F, despite of its simplicity,
    manages to make a difference on top of all other rules.

\section{Bernstein Bases}                                \label{sec:Ber}

    Bernstein bases and Bernstein coefficients for polar codes were
    studied in \cite{DrB20, DrC21, DrB22, DDJ23}.  Bernstein
    coefficients form the closest representation of the shape of a
    function as they are the ``control points'' of a Bézier curve
    \cite{CDr21}.  They are also studied in the context of reliability
    polynomials \cite{ChC98, BCC21}.  In this section, we discuss the
    relation between Bernstein coefficients and the partial order $\>$.

    \begin{definition} [\cite{wBP, DBS11}]
        The Bernstein basis of degree $n$ is the set $\{ \binom{n}{i}
        x^i (1 - x) ^{n-i} : i = 0, \dotsc, n \}$.  It is a basis for
        polynomials of degree at most $n$.
    \end{definition}

    \begin{definition}
        Let $f$ be a polynomial with degree at most $n$.  Let $\Ber{n,
        f}$ be the expansion of $f$ in terms of the Bernstein basis of
        degree $n$: \[ f(x) = \sum_{i=0}^{n} B_i \binom{n}{i} x^i (1 -
        x)^{n-i}. \] $B_i$ are called the \emph{Bernstein coefficients}.
        The collection of rescaled coefficients $N_i \coloneqq B_i
        \binom{n}{i}$ is called the \emph{$N$-form} of $f$.
    \end{definition}

\subsection{Coefficient-wise partial order}

    We define an approximation of $\>$ using Bernstein basis.

    \begin{definition}
        Let $\alpha \in \{0, 1\}^\ell$ and $\gamma \in \{0, 1\}^m$.
        Suppose $n \geq \max(2^\ell, 2^m)$.  We say $\alpha \>_\Ber{n}
        \gamma$ if the coefficients in $\Ber{n, I_\alpha - I_\gamma}$
        are all nonnegative.
    \end{definition}

    \begin{proposition}                             \label{pro:Ber-dual}
        $\alpha \>_\Ber{n} \gamma$ iff $\bar\alpha \<_\Ber{n}
        \bar\gamma$.
    \end{proposition}

    \begin{proposition}                             \label{pro:Ber=>std}
        So long as $n \geq \max(2^\ell, 2^m)$, we have $\alpha
        \>_\Ber{n} \gamma$ implying $\alpha \> \gamma$.
    \end{proposition}

    Proposition~\ref{pro:Ber=>std} indicates that we can use
    $\>_\Ber{n}$ to generate some comparable pairs.  Fig.~\ref{fig:std}
    shows that it almost enumerates all pairs.  This, as we see it, is
    because the converse is almost true.

    \begin{theorem} [\cite{PoR00}]                  \label{thm:std=>Ber}
        \def\footnote#1{}
        Given\footnote{$\Delta \coloneqq I_\alpha - I_\gamma> 0$ is
        stronger than $\Delta \geq 0$ as in the definition of $\alpha \>
        \gamma$; what could go wrong is the presence of $(2x^2 - 1)^2$
        as a factor of $\Delta$} $I_\alpha(x) > I_\gamma(x)$ for all $x
        \in (0, 1)$, we have $\alpha \>_\Ber{n} \gamma$ for sufficiently
        large $n$.
    \end{theorem}

    \begin{IEEEproof}
        This is a direct consequence of the classical results regarding
        whether a positive polynomial over $(0, 1)$ has positive
        Bernstein coefficients.  Bernstein proved that this is indeed
        true if $n$ is made large enough.  Subsequent works addressed
        how large $n$ needs to be.  See \cite[Section~1]{PoR00} for
        discussion.
    \end{IEEEproof}

    There is another (in fact, more effective) approach to prove
    inequalities using Bernstein basis: subdivision.


    \begin{theorem} [\cite{BCR08}] \label{thm:division}
        If $f$ is a degree-$n$ polynomial positive over $[0, 1]$, then
        there are division points $0 = d_0 < \dotsc < d_\ell = 1$ such
        that the Bernstein coefficients of $f$ with respect to each
        subinterval $[d_i, d_{i+1}]$ are all positive.
    \end{theorem}

    In \cite{BCR08}, Theorem~\ref{thm:division} is extended to an
    algorithm that outputs (a) the division points to certify that $f >
    0$, or $f < 0$ or (b) a subinterval $[a, b]$ plus a factor $g \mid
    f$ such that $g(a) g(b) < 0$ to witness that $f$ has roots in $[0,
    1]$.  This is a very reliable method to prove or disprove any
    inequality of the form $\alpha \> \gamma$ in finite time, and is the
    very method used to produce Fig.~\ref{fig:std} and the proof of
    Proposition~\ref{pro:rsE}.

    There is a library worth of literature that studies Bernstein
    coefficients, especially when they play the role of the number of
    size-$i$ subsets in an upward-closed family.  The following
    nontrivial fact, in particular, adds one more reason to why we
    should study them.

    \begin{theorem}
        [Sperner and Kruskal--Katona {\cite[Section~5]{ChC98}}]
        Bernstein coefficients $B_i$ are non-decreasing in $i$.
    \end{theorem}

\subsection{The first nontrivial coefficient}

    \begin{definition}
        Let $f$ be a nonzero polynomial of degree at most $n$.  Let
        $N_i$ be the $N$-form of $\Ber{n, f}$.  The \emph{exponent} of
        $f$ is the smallest index $i$ such that $N_i$ is nonzero.  The
        \emph{mantissa} of $f$ is $N_\text{the exponent of $f$}$.
    \end{definition}

    Notice that $f(\varepsilon) = \text{mantissa} \cdot \varepsilon
    ^\text{exponent} (1 + O(\varepsilon))$ when $\varepsilon \to 0$.

    \begin{theorem} [{\cite[Theorem~3]{DrB20}}]
                                           \label{thm:exponent-mantissa}
        Let $\alpha \in \{0, 1\}^\ell$.  Let $n \geq 2^\ell$.  Let $N_i$
        be the $N$-form of $\Ber{n, I_\alpha}$.  Then the exponent of
        $I_\alpha$ is $2^z$, where $z$ is the number of zeros in
        $\alpha$; moreover, the mantissa of $I_\alpha$ is such that
        \[
            \log_2 (N_{2^z}) = \sum_{k=1}^{\ell-z} 2^
            \text{
                number of $0$'s to the right
                of the $k$th $1$ in $\alpha$
            }.
        \]
    \end{theorem}

    \begin{IEEEproof} [Sketch]
        Note that $I_0(x) = x^2$ and $I_1(x) = 2x + O(x^2)$.  So $I_0$
        doubles the exponent and $I_1$ maintains it.  Moreover, $I_0$
        squares the mantissa and $I_1$ doubles it.
    \end{IEEEproof}
    
    Studying exponents and mantissas gives us the leverage to determine
    whether $W^\alpha$ or $W^\gamma$ is better when the underlying $W$
    is very reliable.  This is the theme of \cite[Section~IV-C]{DrB20}.
    On the other hand, it can also be used to generate necessary
    conditions for the comparability of $\alpha$ and $\gamma$.  The
    following proposition gives an easy class of incomparable pairs.

    \begin{proposition}                               \label{pro:more01}
        If $\alpha$ has more ones than $\gamma$ does and $\alpha$ has
        more zeros than $\gamma$ does, then $\alpha$ and $\gamma$ are
        incomparable by $\>$, i.e., $\alpha \not\> \gamma$ and $\alpha
        \not\< \gamma$.
    \end{proposition}

    Proposition~\ref{pro:more01} is not useful if one wants to compare
    $\alpha$ and $\gamma$ having the same length.  So in the next
    subsection we are going to craft some examples leveraging the
    knowledge of mantissas.

\subsection{Use knowledge of coefficients to construct incomparables}
                                                       \label{sec:incom}

    \begin{definition}
        We write $\alpha \geq_\ato \gamma$ if $I_\alpha - I_\gamma = 0$
        or the mantissa of $I_\alpha - I_\gamma$ is positive.  We write
        $\alpha \sim_\ato \gamma$ if $I_\alpha$ and $I_\gamma$ share the
        same exponent and mantissa.  We write $\alpha >_\ato \gamma$ if
        $I_\alpha$ has a smaller exponent than $I_\gamma$ does, or they
        have the same exponent but the former has a higher mantissa.  
    \end{definition}

    $\geq_\ato$, $\sim_\ato$, and $>_\ato$ imply that $\lim_{x\to0}
    I_\alpha(x) / I_\gamma(x)$ is $\geq 0$, $= 1$, or $> 1$,
    respectively.  There are also ``$\atl$'' versions, which ask for the
    same criteria at the neighborhood of $x = 1$.

    \begin{definition}
        We write $\alpha \geq_\atl \gamma$ if $\bar\alpha \leq_\ato
        \bar\gamma$.  We write $\alpha \sim_\atl \gamma$ if $\bar\alpha
        \sim_\ato \bar\gamma$.  We write $\alpha >_\atl \gamma$ if
        $\bar\alpha <_\ato \bar\gamma$.
    \end{definition}

    \begin{proposition}                              \label{pro:std=>at}
        If $\alpha \> \gamma$ then $\alpha \geq_\ato \gamma$ and $\alpha
        \geq_\atl \gamma$.
    \end{proposition}

    Example usage of Proposition~\ref{pro:std=>at}: $100001$ and
    $011000$ are incomparable under $\>$ because $100001 >_\ato 011000$
    and $100001 <_\atl 011000$.

    \begin{proposition}                               \label{pro:at-con}
        If $\alpha >_\ato \gamma$ and $\kappa \sim_\ato \lambda$, then
        $\alpha \kappa \geq_\ato \gamma \lambda$ and $\kappa \alpha
        \geq_\ato \lambda \gamma$.
    \end{proposition}

    Constructing more incomparable pairs becomes easy thanks to
    Proposition~\ref{pro:at-con}.  For instance, $(\kappa 100001
    \lambda, \kappa 011000 \lambda)$ is an incomparable pair for
    arbitrary $\kappa$ and $\lambda$.

\begin{figure*}
    \centering
    \begin{tikzpicture} [x=11mm, y=3mm]
        \def\pgfplotdrawwithnode (#1) (#2) [#3] #4;{
            \draw (#1) rectangle node [scale=2^((1-#3)/4)] {#4} (#2);
        }
        \def\pgfplotdrawrectangle (#1) (#2);{
            \draw (#1) rectangle (#2);
        }
        \input{influence8.tex}
    \end{tikzpicture}
    \caption{
        How to select beta?  The height of a rectangle labeled $\alpha$
        is $I_{\alpha0}^{-1} (1/2) - I_{\alpha1}^{-1} (1/2)$.  This
        number is the \emph{influence} of the last digit on the halfway
        point.  The choice $\bb = 2^{1/4}$ represents an estimate that
        the average height of the rectangles on level $\ell$ is about
        $2^{-\ell/4}$.  That being said, (a) $\bb = 2^{1/3.6}$ seems to
        be a better choice, at least for BECs; (b) since the influence
        depends on $\alpha$, neither $2^{1/4}$ nor $2^{1/3.6}$ provides
        rigorous ordering.
    }                                                  \label{fig:scale}
\end{figure*}

\begin{figure*}
    \centering
    \begin{tikzpicture} [x=0.64mm, y=3mm]
        \def\pgfplotdrawposition(#1,#2,#3,#4)#5;{
            \draw [every node/.style={inner sep=0,scale=1/6,fill=white}]
                (#3, 3) node {#5\vphantom{0}} --
                (#1, 2) node {#5\vphantom{0}} --
                (#4, 1) node {#5\vphantom{0}} --
                (#2, 0) node {#5\vphantom{0}}
                (#1, -1) node {#5\vphantom{0}} --
                (#2, -2) node {#5\vphantom{0}} --
                (#3, -3) node {#5\vphantom{0}} --
                (#4, -4) node {#5\vphantom{0}}
            ;
        }
        \input{permutation8.tex}
        \draw [every node/.style={scale=2/3}]
            (0, 3) node [left] {$\avg$}
            (0, 2) node [left] {$\bb = 2^{1/4}$}
            (0, 1) node [left] {$\hlf$}
            (0, 0) node [left] {$\bb = 2^{\frac{1}{3.627}}$}
            (0, -1) node [left] {$\bb = 2^{1/4}$}
            (0, -2) node [left] {$\bb = 2^{\frac{1}{3.627}}$}
            (0, -3) node [left] {$\avg$}
            (0, -4) node [left] {$\hlf$}
            (255, 2.5) node [right] {492}
            (255, 1.5) node [right] {582}
            (255, 0.5) node [right] {285}
            (255, -1.5) node [right] {313}
            (255, -2.5) node [right] {195}
            (255, -3.5) node [right] {110}
        ;
    \end{tikzpicture}
    \caption{
        Matchings among four total orders on $\{0, 1\}^8$.  The string
        $11110000$ is shorthanded as f0.  The Kendall tau distances (the
        numbers of crossings) are marked on the right.
    }                                                  \label{fig:total}
\end{figure*}

\section{Beta Expansion}                                \label{sec:beta}

    Beta expansion \cite{HBL17} was defined not to partial-order the
    channels, but to give, for each block length, a totally-ordered list
    so a user will simply take the first $k$ strings to construct a code
    of dimension $k$.  This is essentially a curve-fitting problem.  In
    \cite{ZZL18} the authors models this with more parameters.

    In this paper, we discuss whether beta expansion is compatible with
    the rule sets covered in Section~\ref{sec:rs}.   We also try to
    ``explain'' beta expansion using scaling exponent.

    \begin{definition}
        Let $\bb$ be a positive real number.  The beta expansion of any
        string $\alpha = a_1 \dotsm a_\ell \in \{0, 1\} ^\ell$ is
        defined as
        \[ \alpha_\bb \coloneqq \sum_{i=1}^{\ell} a_i \bb^{\ell-i}. \]
    \end{definition}

    \begin{definition}
        We say $\alpha \geq_\bb \gamma$ if $\alpha_\bb \geq \gamma_\bb$
        for the beta value understood from the context.
    \end{definition}

\subsection{Compatibility of beta and rule sets}

    \begin{proposition}                          \label{pro:beta-concat}
        If $\alpha \geq_\bb \gamma $ and $\kappa \geq_\bb \lambda$ while
        $\kappa$ and $\lambda$ sharing the same length, then $\alpha
        \kappa \geq_\bb \gamma \lambda$.
    \end{proposition}

    \begin{proposition} [{\cite[Proposition~3]{HBL17}}]
                                                   \label{pro:rsA=>beta}
        $1 \geq_\bb 0$ if $\bb \geq 0$.
    \end{proposition}

    \begin{proposition} [{\cite[Proposition~3]{HBL17}}]
                                                   \label{pro:rsB=>beta}
        $10 \geq_\bb 01$ if $\bb \geq 1$.
    \end{proposition}

    \begin{proposition} [{\cite[Proposition~1]{HBL17}}]
                                                   \label{pro:beta-dual}
        For $\alpha$ and $\gamma$ of the same length,
        $\alpha \geq_\bb \gamma$ iff $\bar\alpha \leq_\bb \bar\gamma$.
    \end{proposition}

    \begin{proposition}                            \label{pro:rsC=>beta}
        $10 0^k 01 \geq_\bb 01 0^k 10$ for all $k$ if $\bb \geq 1$.
    \end{proposition}

    \begin{IEEEproof}
        $(10 0^k 01)_\bb - (01 0^k 10)_\bb = (\bb - 1) (\bb^{k+2} - 1)$,
        which is nonnegative as $\bb \geq 1$.
    \end{IEEEproof}

    \begin{proposition}                            \label{pro:rsD-beta}
        $0^k 1^{2^k} \geq_\bb 1^k 0^{2^k}$ implies $\bb \leq 2^{-1/k}$.
        So beta expansion and RS-D are not compatible due to
        $\lim_{k\to\infty} 2^{-1/k}
        = 1$.
    \end{proposition}

    \begin{IEEEproof}
        $1 \leq (0^k 1^{2^k})_\bb / (1^k 0^{2^k})_\bb \leq \bb^{-k} +
        \bb^{-2k} + \bb^{-3k} + \dotsb = \bb^{-k} / (1 - \bb^{-k})$.
        This implies $\bb^{-k} \geq 1/2$.
    \end{IEEEproof}

    \begin{proposition}                            \label{pro:rsE=>beta}
        $\tau_1 \dotsm \tau_{2^k} \leq_\bb \overline{\tau_1 \dotsm
        \tau_{2^k}}$ if $\bb \geq 1$.
    \end{proposition}

    \begin{IEEEproof}
        $(\overline{\tau_1 \dotsm \tau_{2^k}})_\bb - (\tau_1 \dotsm
        \tau_{2^k})_\bb = (\bb^{2^0} - 1) \* (\bb^{2^1} - 1) \* \dotsm
        \* (\bb^{2^{k-1}} - 1)$, which is nonnegative as $\bb \geq 1$.
    \end{IEEEproof}

\subsection{A theoretical connection between beta and scaling}
                                                       \label{sec:scale}

    Scaling exponent is the answer to the following question: at block
    length $2^\ell$, how many synthetic channels ``stay mediocre''?
    That is, how many $\gamma \in \{0, 1\}^m$ are such that $\varepsilon
    < H(W^\gamma) < 1 - \varepsilon$.  As it turns out, this number is
    about $2^{m - m/4}$ for AWGN channels and $2^{m - m/3.627}$ for BECs
    \cite{KMT10, HAU14}.

    In beta expansion, the $i$th bit of $\alpha$ is endowed with an
    ``influence'' of strength $\bb^{\ell-i}$ (see also
    Fig.~\ref{fig:scale}).  But, it is very clear that $a_i$ is
    influential only if $W^{a_1 \dotsm a_{i-1}}$ is mediocre.  Hence, a
    hand-waving argument is that $\bb^{\ell-i}$ should be in proportion
    to the fraction of mediocre channels, which is $2^{-i/4}$ for AWGN
    channels.  This suggests $\bb \approx 2^{1/4}$, coinciding with the
    value recommended in \cite{HBL17}.

    To prove our point, we turn to BECs.  Since the fraction of mediocre
    channels is now $2^{-i/3.627}$, we hypothesize that $\bb =
    2^{1/3.627}$ should be ``more suitable'' than $\bb = 2^{1/4}$.  We
    measure the Kendall tau distance \cite{wKTD} among four total
    orders: $\geq_\bb$ using $\bb = 2^{1/4}$ or $\bb = 2^{1/3.627}$, and
    $\geq_\avg$ and $\geq_\hlf$ defined in Section~\ref{sec:cut}.  We
    see in Fig.~\ref{fig:total} that $2^{1/3.627}$, $\avg$, and $\hlf$
    lead to similar total orders while $\bb = 2^{1/4}$ leads to an
    outlier.  This indicates that it is appropriate to sync $\bb$ with
    the scaling exponent.

\section{Threshold---Area or Halfway Point?}             \label{sec:cut}

    Inspired by beta expansion, we define total orders so that the first
    $k$ strings will likely form a good polar code.  Or, a more
    conservative approach is to include the first $k/2$ strings and
    carefully examine the next $k$ strings (cf.\ \cite{ZGS18}).  For
    either goal, a representative total order is in need.

\subsection{Average}

    \begin{definition} [{\cite[Definition~10]{DrC21}}]
        Define $\avg(\alpha)$ to be $\int_0^1 I_\alpha(x) \,dx$.  We
        write $\alpha \geq_\avg \gamma$ when $\avg(\alpha) \geq
        \avg(\gamma)$.
    \end{definition}

    \begin{proposition} [{\cite[Proposition~7]{DrC21}}]
                                                    \label{pro:std=>avg}
        $\alpha \geq_\avg \gamma$ if $\alpha \> \gamma$.
    \end{proposition}

    \begin{proposition} [{\cite[Lemma~5]{DrC21}}]   \label{pro:avg-dual}
        $\alpha \geq_\avg \gamma$ iff $\bar\alpha \leq_\avg \bar\gamma$.
    \end{proposition}
    
    \begin{proposition}                           \label{pro:martingale}
        $\avg(\alpha0) + \avg(\alpha1) = 2\avg(\alpha)$.
    \end{proposition}

    Proposition~\ref{pro:martingale} implies that, if $A_1, A_2, \dotsc$
    are iid random bits, then $\avg(A_1 \dotsm A_t)$ is a martingale as
    $t$ increases.  This martingale is bounded and hence converging.
    The limit is the $\vartheta$ function in \cite[Proposition~3]{Gei18}
    and the $z^*$ function in \cite[Lemma~11]{HAU14}.  Hence $\avg(A_1
    \dotsm A_t)$ can also be seen as the Doob martingale of $z^*$.

\subsection{Halfway point}

    \begin{definition}
        Define $\hlf(\alpha) \coloneqq I_\alpha^{-1}(1/2)$.  We say
        $\alpha \geq_\hlf \gamma$ if $\hlf(\alpha) \leq \hlf(\gamma)$.
    \end{definition}
    
    \begin{proposition}                             \label{pro:std=>hlf}
        $\alpha \> \gamma$ implies $\alpha \geq_\hlf \gamma$.
    \end{proposition}

    \begin{proposition}                             \label{pro:hlf-dual}
        $\alpha \geq_\hlf \gamma$ iff $\bar\alpha \leq_\hlf \bar\gamma$.
    \end{proposition}

    The halfway point was studied in \cite{Gei18, WuS19}.  It and $\avg$
    are eventually the same function as the length of string goes to
    infinity because the reliability polynomials experience hard
    thresholds \cite{OrR19}.  Halfway point outshines average because it
    is as easy to compute as beta expansion (cf.\ \cite{DDJ23}).

\subsection{A Fast preorder}                            \label{sec:fast}

    We conclude this paper with an intersection of three preorders
    inspired by Section~\ref{sec:incom}.

    \begin{definition}
        We say $\alpha \>_\fst \gamma$ if all three bullets are met:
        \begin{itemize}
            \item $\alpha >_\ato \gamma$ or $\alpha \sim_\ato \gamma$;
            \item $\alpha >_\atl \gamma$ or $\alpha \sim_\atl \gamma$;
            \item $\alpha \geq_\hlf \gamma$.
        \end{itemize}
    \end{definition}

    \begin{proposition}                             \label{pro:std=>fst}
        $\alpha \> \gamma$ implies $\alpha \>_\fst \gamma$.
    \end{proposition}
    
    \begin{proposition}                             \label{pro:fst-dual}
        $\alpha \>_\fst \gamma$ iff $\bar\alpha \<_\fst \bar\gamma$.
    \end{proposition}

    $\>_\fst$ is easy to compute as $>_\ato, \sim_\ato, >_\atl,
    \sim_\atl$ involve only counting $0$ and $1$ and $\geq_\hlf$ is a
    matter of chaining $I_0^{-1}$ and $I_1^{-1}$.  The fast partial
    order is empirically a very successful approximation of $\>$.  See
    Fig.~\ref{fig:std} for how unnoticeable the difference is.

\section{Acknowledgements}

    V.-F. Dr\u{a}goi was financed by the Romanian Ministry of Education
    and Research, CNCS-UEFISCDI, Grant Number:
    PN-III-P4-ID-PCE-2020-2495.

\IEEEtriggeratref{22}
\bibliographystyle{IEEEtran}
\bibliography{Polder-Finer-25.bib}

\end{document}